\documentclass[aps,prb,amsmath,amssymb,twocolumn]{revtex4-2}

\usepackage{graphics}%
\usepackage{dcolumn}%
\usepackage{tikz}
\usepackage{bm}%
\usepackage{amsmath}
\usepackage{bbold}%

\usepackage{hyperref}%

\DeclareMathOperator{\Tr}{Tr}
    
\graphicspath{{figures/}}

\begin{document}

\preprint{APS/123-QED}

\title{Emergence of the 2nd Law in an Exactly Solvable Model of a Quantum Wire}%

\author{Marco A. Jimenez-Valencia}
\email{marcojv@arizona.edu}
\affiliation{%
 Department of Physics, University of Arizona, 1118 E.\ 4th Street, Tucson, Arizona 85721 
}%

\author{Charles A. Stafford}%
\affiliation{%
 Department of Physics, University of Arizona, 1118 E.\ 4th Street, Tucson, Arizona 85721 
}%

\date{\today}%
                                      
\begin{abstract}

As remarked by Boltzmann, the Second Law of Thermodynamics is notable for the fact that it is readily proved using elementary statistical arguments, but becomes harder and harder to verify the more precise the microscopic description of a system.  In this article, we investigate one particular realization of the 2nd Law, namely Joule heating in a wire under electrical bias.
We analyze the production of entropy in an exactly solvable model of a quantum wire wherein the conserved flow of entropy under unitary quantum evolution is taken into account using an exact formula for the entropy current of a system of independent quantum particles.  In this exact microscopic description of the quantum dynamics, the entropy production due to Joule heating does not arise automatically.  Instead, we show that the expected entropy production is realized in the limit of a large number of local measurements by a series of floating thermoelectric probes along the length of the wire, which inject entropy into the system as a result of the information obtained via their continuous measurements of the system.  The decoherence resulting from inelastic processes introduced by the local measurements is essential to the phenomenon of entropy production due to Joule heating, and would be expected to arise due to inelastic scattering in real systems of interacting particles.

\end{abstract}

\maketitle

\section{Introduction}

Deriving the 2nd Law of Thermodynamics---the law of increase of entropy---from microscopic dynamics is a project dating back to Boltzmann.  Whether using classical or quantum dynamics, the issue is that entropy is conserved under exact time evolution, taking all degrees of freedom into account.  The 2nd Law arises because, in practice, a macrosopic observer is unable to keep track of all the information encoded in the microscopic degrees of freedom of a system, so that the information \cite{Shannon} needed to specify those degrees of freedom (and hence the system entropy) at later times continuously increases \cite{landauerElectricalTransportOpen1987}.  This scenario has been implemented for open quantum systems to describe the increase of the von Neumann entropy \cite{vonneumann} associated with the reduced density matrix of a quantum system coupled to a macroscopic reservoir \cite{Esposito2010}, and has led to a large body of work on entropy production in the quantum thermodynamics of such systems \cite{Landi2021}. A contrary viewpoint concerning the role of information and irreversibility in the thermodynamics of quantum systems is discussed in \cite{goldsteinGibbsBoltzmannEntropy2019}.

However, the phenomenon of {\em Joule heating}, the key manifestation of the 2nd Law of Thermodynamics in electrical transport, which has been observed experimentally even in atomic-scale conductors \cite{Lee2013,mengesTemperatureMappingOperating2016a,Reddy2017,Menges2017}, %
has not yet been derived from a fully microscopic quantum Hamiltonian (although phenomenologial approaches have been presented for instance in Refs.\ \cite{Bringuier24,koizumi2025}), and is apparently not amenable to the reduced-state approach pioneered in Ref.\ \cite{Esposito2010} and reviewed in Ref.\ \cite{Landi2021}.  Computational approaches such as Ref.\ \cite{PopMonteCarlo} touch upon the electron-phonon relaxation mechanisms leading to Joule heating, reminiscent of the reduced-state approach.
A quasi-microscopic description is provided by the two-temperature model \cite{AllenTs, DasSarmaTs, WaldeckerTs, SadasivamTs, SjaksteTs,  MinamitaniTs}, wherein hot electrons with short relaxation times scatter inelastically on longer timescales with phonons of lower temperature that subsequently thermalize to an environment where this temperature can be measured. 

In this article, we analyze the flow of entropy in a quantum electric circuit using a new unitary entropy current formula \cite{evers}, which correctly accounts for the fact that the entropy of the universe, including all microscopic degrees of freedom, is conserved under the electrical transport process.  In this microscopic description of entropy flow, inelastic scattering processes leading to thermalization and entropy production need to be explicitly included in the model in order to account for the 2nd Law.  This is achieved by attaching a large number of floating thermoelectric probes \cite{Bergfield2014,Shastry2016} to a quantum wire, that continuously measure the local temperature and chemical potential along the length of the wire, serving as sources of inelastic scattering and decoherence \cite{buttiker88}.  The information about the local nonequilibrium electron distribution in the wire obtained by these probes is not stored, but is injected as entropy into the quantum wire.

The physical picture that describes these measurements is the following: When a steady flow propagates into a reservoir, the relaxation processes that it undergoes lead to entropy production. Eventually, the distribution describing this flow into a reservoir relaxes until it is a linear perturbation of the equilibrium distribution at infinity. In other words, macroscopic measurements in electric circuits imply that the particles involved in the transport have undergone processes of relaxation into equilibrium with the leads they end up at, generating entropy. Examples of protocols for direct measurement of entropy in these types of systems can be found in \cite{hartmanDirectEntropyMeasurement2018,childEntropyMeasurementStrongly2022,kleeorinHowMeasureEntropy2019}. These processes occur naturally in real circuits due to the fact that there are multiple sources of decoherence that in turn permit this relaxation. This implies that only energy and particle fluxes are conserved, a situation that is described by the conventional formula for heat currents, from which a description of entropy currents can be defined for quasi-reversible processes in accordance with thermodynamics by dividing heat currents by temperature $T$. It is an open question how to define entropy currents in general irreversible processes \cite{Strunk2021}.

The unitary description of entropy flow utilized in the present article expands on the analyses of Refs.\ \cite{marco,evers}, which investigate persistent and microscopic flows of entropy in open quantum systems, and is to be contrasted with the conventional quantum formulas \cite{buttiker1,Imry1,bergnano2009} describing the flow of charge and heat into macroscopic reservoirs, that implicitly assume complete thermalization of the outgoing distributions at infinity.

This paper is organized in the following way: A description of both the paradigm and the proposed (unitary) approaches for describing entropy currents and the differences in their consequences in terms of entropy generation is discussed in \ref{sec:discrep}. In section \ref{sec:emergence}, following an explanation of these discrepancies, a proposal to recover the conventional entropy generation from the unitary description is tested with positive results in an infinite-chain system. The limit behavior of the disparity in the different approaches is explained as an end effect in \ref{sec: 1_N} and a summary of the results can be found in \ref{sec:summary}. Three appendices \ref{apx:green}, \ref{apx:singlelimit} and \ref{apx:oneandtwo} describe the basics of the NEGF formalism used in this work, the limit achievable by a single probe in the system in terms of entropy generation, and the quasi-classical resistance-network behavior in the chain as well as its dependence to the system parameters, respectively.

\section{Discrepancies between unitary and dissipative entropy flows} \label{sec:discrep}

The introduction of unitary entropy currents brings about the challenge of reconciling it with previous descriptions of the flow of entropy and heat. Among these is the discussion of entropy production. 

The conventional description of (dissipative) heat flow in a multiterminal system is given by the B\"uttiker-Sivan-Imry formula \cite{buttiker1,Imry1,stoneWhatMeasuredWhen1988,bergnano2009}
\begin{equation}
I^{(\nu)}_\alpha=\frac{1}{h}\int d\epsilon \left(\epsilon-\mu_\alpha\right)^\nu\sum_\beta T_{\alpha\beta}\left[f_\beta-f_\alpha\right], \label{BSI}
\end{equation}
where $\alpha$ labels the reservoirs, $h$ is Planck's constant, $\nu=0$ corresponds to particle current and $\nu=1$ to heat current. $T_{\alpha\beta}(\epsilon)$ is the transmission function from reservoir $\beta$ to $\alpha$ and $f_\alpha(\epsilon)=\left(e^{\beta_\alpha
	(\epsilon-\mu_\alpha)}+1\right)^{-1}$ is the Fermi-Dirac distribution function of reservoir $\alpha$, where $\beta_\alpha=1/k_B T_\alpha$.

Eq.\ \eqref{BSI} implies that the electrical work supplied as electrons are transmitted between the reservoirs is converted entirely into heat, a process known as Joule heating:
\begin{equation}
    \sum_\alpha I_\alpha^{(1)} = -\sum_\alpha \mu_\alpha I_\alpha^{(0)}, \label{jouleh}
\end{equation}
where the lhs is the total rate of heat production and the rhs is the electrochemical power ${\cal P}$ supplied by the external electric bias.
The total entropy production rate due to Joule heating is
\begin{equation}
    \sum_\alpha \frac{I_\alpha^{(1)}}{T_\alpha} \stackrel{T_{\alpha}=T_0}{=} \frac{1}{T_0} \left( -\sum_\alpha \mu_\alpha I_\alpha^{(0)} \right) \geq 0,
    \label{eq:joules}
\end{equation}
where positivity is ensured since particles %
flow spontaneously from high to low chemical potential.
The conventional expression successfully  describes the entropy production in the irreversible process of  currents flowing between macroscopic reservoirs.

In contrast with this, the unitary flow of entropy governed by the Schr\"odinger equation leads to the conserved entropy current \cite{jimenez25c}
\begin{equation}
     I^{S}_{\alpha}=\frac{1}{h}\int d\epsilon \sum_\beta T_{{\alpha}\beta}\left[s_\beta\left(\epsilon\right)-s_\alpha\left( \epsilon\right)\right] \label{unitarys},
\end{equation}
where $s_{\alpha}(\epsilon)$ is the single-particle entropy contribution from a distribution $f_\alpha$ at energy $\epsilon$. In the case of Fermions $s_{\alpha}(\epsilon) = -k_B \left[ f_\alpha\ln f_\alpha + (1-f_\alpha)\ln (1-f_\alpha) \right]$ \footnote{This can be substituted by a Bosonic entropy $s^b(\epsilon) = - k_B \left[ f^b\ln f^b - (1+f^b)\ln (1+f^b) \right]$, where $f^b(\epsilon)= \left(e^{\beta(\epsilon-\mu)}-1\right)^{-1}$ to describe for instance phonon or photon entropy fluxes if chemical potential is zero or conserved bosonic particles otherwise.}.
Eq.\ \eqref{unitarys} is analogous to the multi-terminal B\"uttiker formula \cite{buttiker1} for charge transport, and is justified within a local treatment of entropy flow \cite{marco,evers} that keeps track of all microscopic degrees of freedom.
A similar formula was derived in Ref.\ \cite{strunkQuantumTransportParticles2021}.
The key point to highlight in this context is that this expression takes into account a higher degree of \textit{knowledge} of the system with respect to the previous case, by respecting and accounting for the exact unitary quantum evolution of all degrees of freedom in the system and reservoirs.

A central consequence of Eq.\ \eqref{unitarys} is that the total entropy is conserved
\begin{equation}
    \sum_\alpha I_\alpha^S = 0. \label{noSprod}
\end{equation}
It is important to note that this should not be unexpected if this formula is understood, the same way it is clear that the unitary evolution of a quantum system leads to an unchanging von Neumann entropy. To describe the currents via Eq.\ \eqref{unitarys} is to fully take into account degrees of freedom that are disregarded in a dissipative framework, one that assumes decoherence and thermalization of particles that come from a reservoir and end up in another one, without modeling its mechanisms explicitly.  This assumption of decoherence is implicit \cite{stoneWhatMeasuredWhen1988} in the conventional current formulas \eqref{BSI} %
wherein an order of limits \cite{baranger93,baranger89} is taken to prevent coherent backscattering in a noninteracting translationally-invariant reservoir (see also Refs.\ \cite{corneanRigorousProofLandauer2005,corneanAdiabaticallySwitchedelectricalBias2008,corneanMathematicalAccountNEGF2018} for modern mathematically rigorous exploration of this issue).

This discrepancy in the description of the production of entropy between the conventional dissipative entropy current and the exact unitary entropy current formulas is the main subject of this work. 

\section{Emergence of entropy generation: unitary to dissipative} \label{sec:emergence}

If Eq.\ \eqref{unitarys} successfully describes the flow of entropy at the microscopic level, then there must be a way to recover the expression for the entropy production due to Joule heating, Eq.\ \eqref{eq:joules}, by leaving out the information contained in the degrees of freedom that are ignored in the conventional dissipative current formula, Eq.\ \eqref{BSI}. 
A key hypotheses of this work is that this discrepancy %
can be traced back to the mechanisms of decoherence and inelastic scattering, %
which are implicitly assumed in the derivation of Eq.\ \eqref{BSI} to lead to complete thermalization of the outgoing fluxes into the macroscopic reservoirs.

\subsection{Entropy production in an infinite chain}

\begin{figure}[h]
\includegraphics[width=1\linewidth]{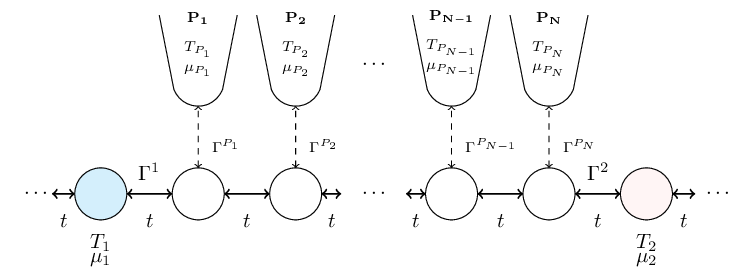}
	\caption{ Model system consisting of an infinite %
    tight-binding chain coupled to $N$ floating thermoelectric probes, serving as sources of decoherence and thermalization.  
    Information acquired through the probes' continuous measurements results in entropy injected into the wire.
    The semi-infinite sections to the left and right of the probe region represent the source and drain reservoirs. \label{fig:model}  }
\end{figure}

As a tractable model to study the hypothesized emergence of irreversibility and entropy generation from the unitary entropy current formula \eqref{unitarys}, we consider a quantum wire modeled as an infinite tight-binding chain connecting source and drain reservoirs, coupled to $N$ floating thermoelectric probes whose continuous measurements \cite{Bergfield2014,Shastry2016} of the local temperature and chemical potential serve as sources of decoherence and inelastic scattering in the wire (see Fig.\ \ref{fig:model}).  
The source and drain reservoirs are held at a common temperature $T_0$ but with an electric bias $\mu_1-\mu_2$ between their chemical potentials. %

The total Hamiltonian is 
\begin{equation}
    H = H_S + H_R + H_P + H_{\text{coupling}},
\end{equation}
where %
\begin{equation}
    H_S = t\sum_{i=1}^{N-1} d^\dagger_i d_{i+1} + \mbox{h.c.} \label{sysh}
\end{equation}
describes the central region of the wire, $t$ being the tight-binding hopping matrix element between neighboring sites of the chain, the
reservoirs are described by
\begin{equation}
    H_R = \sum_{\alpha = a,b} t \left( \sum_{j=1}^\infty c_{j,\alpha}^\dagger c_{j+1,\alpha} +\mbox{h.c}. \right), \label{resh} 
\end{equation}
where $\alpha$ labels the reservoirs, $a$ for the source on the left and $b$ for the drain on the right. The labeling is such that the subscript $(1,a)$ corresponds to the site at the interface between the left reservoir and the first site in the system, and $(1,b)$ to the site at the interface between the right reservoir and the $N$th site in the system counted from the first probe. The floating probes are also reservoirs described as metallic Fermi gases
\begin{equation}
    H_{P}=\sum_{\beta=1}^N H_{P_\beta}=\sum_{\beta=1}^{N}\,\sum_{k\in P_\beta} \varepsilon_{k\beta} c_{k,P_\beta}^\dagger c_{k,P_\beta} \label{probeh},
\end{equation}
and the couplings are described by
\begin{equation}
    H_{\text{coupling}}=t \left(c_{1,a}^\dagger d_1 + c_{1,b}^\dagger d_N  \right) + \sum_{\beta=1}^N\sum_{k\in P_\beta} V_{k\beta}c_{k,P_\beta}^\dagger d_\beta +\mbox{h.c.} \label{couplingsh},
\end{equation}
where the first term corresponds to the coupling of the semi-infinite chains to the left and right and the second term corresponds to the coupling of the $N$ probes to the corresponding sites in the system.

The semi-infinite source and drain reservoirs coupled to the ends of the $N$-site quantum wire can be treated by the NEGF technique, as discussed in Ref.\ \cite{parthprl}, and
the coupling of the probes to the $N$ sites within the quantum wire is treated in the broadband limit, with a constant coupling $\gamma_p$.

\subsection{Floating probe measurements}
Each floating probe labeled by $P_n$ performs a continuous measurement %
that takes pure states in the wire and returns mixed states, acquiring information about the electronic states in the wire and disposing of that information by injecting entropy into the system, not storing it. These probes serve as neither sources of particles nor energy in the wire, and their chemical potentials and temperatures adjust to enforce this floating condition
\cite{buttiker89,buttiker88,buttiker2007,Bergfield2014,Shastry2016} 
\begin{equation}
    I_{P_n}^{(0)} = 0 \\ \text{ and }\  I_{P_n}^{E} = 0,\  \forall\ n=1,\dots, N ,
\end{equation}
where $I_{P_n}^{E}$ is the current of energy into $P_n$, which are equivalent to
\begin{multline}
    I_{P_n}^{(\nu)}(\mu_{P_1},\dots,\mu_{P_N},T_{P_1},\dots,T_{P_N}) = 0, \\ \forall\ n=1,\dots, N, \text{ and } \ \nu=0,1, \label{probecond} 
\end{multline}
where the currents defined in Eq.\ \eqref{BSI} are calculated using NEGF \cite{datta,stefanucci}.

The floating conditions define a set of $2N$ nonlinear equations in $2N$ unknowns, which are the chemical potentials and temperatures of the thermoelectric probes. %
This set of equations is intractable analytically and expensive to solve numerically, as it involves not only nonlinear equations but integrations as well.

An important simplification can be made by 
performing a Sommerfeld expansion of Eq.\ \eqref{probecond}, applicable when $\mu_\alpha \gg k_B T_0$ $\forall \alpha$ as done in \cite{shastry2015}, leading to
\begin{equation}
    I_{P_n}^{(0)} = \frac{1}{h} \sum_\alpha T_{{P_n}\alpha}( \mu_0 )\left(\mu_\alpha-\mu_{P_n}\right)=0 \label{i0ps},
\end{equation}
\begin{multline}
     I_{P_n}^{(1)} = \frac{1}{h} \sum_\alpha T_{{P_n}\alpha}\left( \mu_0 \right)\\ \times\left[ \frac{\left(\mu_\alpha-\mu_{P_n} \right)^2}{2}- \frac{\pi^2k_B^2}{6} \left( T_\alpha^2-T_{P_n}^2 \right)\right]=0 \label{i1ps},    
\end{multline}
where $\mu_0 = \frac{1}{2}\left(\mu_1+\mu_2\right)$ and terms involving $T_{\alpha\beta}'(\mu_0)$ and higher derivatives have been dropped, valid to leading order in the electrical bias, as needed to correctly describe Joule heating. 

The $N$ Eqs.\ \eqref{i0ps} define a linear system for the $N$ unknowns $\mu_{P_n}$, which can then be introduced as parameters for the $N$ Eqs.\ \eqref{i1ps} which define another linear system on the $N$ temperatures of the probes squared $T_{P_n}^2$. This method then not only the $2N$ system of equations into two $N$ systems, but linearizes them and avoids the necessity of integration. %

It is important to highlight that while the currents are calculated in the Sommerfeld expansion, the transmission functions appearing in Eqs.\ \eqref{i0ps} and \eqref{i1ps} are calculated exactly using NEGF.

\begin{figure}[h]	
    \includegraphics[width=1\linewidth]{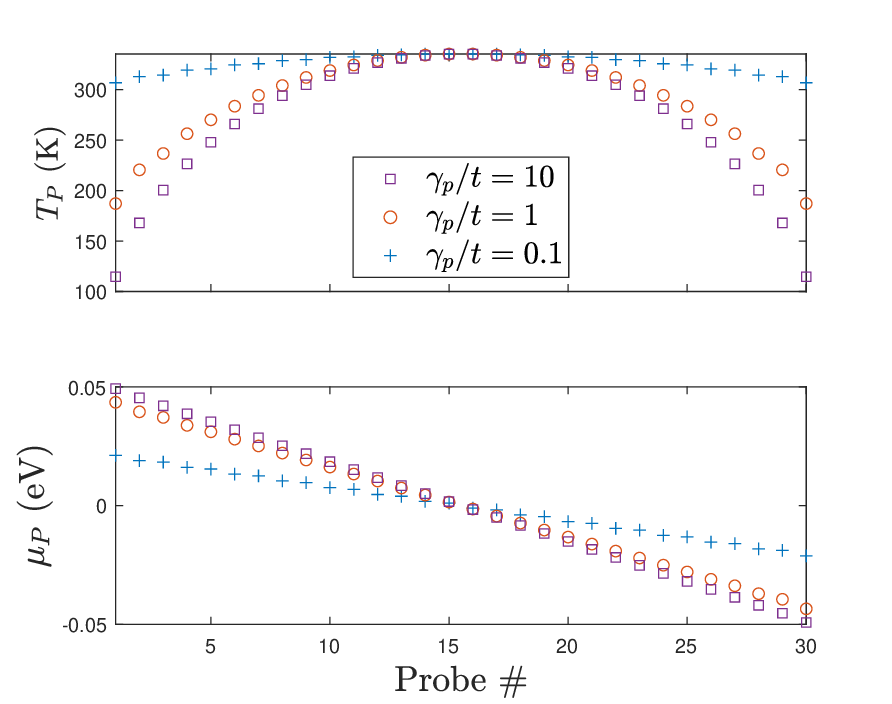}
	\caption{Temperatures (top panel) measured simultaneously with chemical potentials (bottom panel) in an infinite chain with 30 floating probes. %
    Here $T_0=100$K, $t=2.7$eV, $\Delta \mu= 0.1$ eV.}\label{fig:mus1}  
\end{figure}

Fig.\ \ref{fig:mus1} shows chemical potential and temperature profiles along the quantum wire, illustrating the generic behavior found for all but the weakest probe coupling (In the cases of very low $\gamma_p/t$, 2$k_F$ oscillations of both the chemical potential and temperature were observed, as discussed in Refs.\  \cite{buttiker89,demon2013}), where the chemical potential drops nearly linearly between the source and drain contacts, %
and the temperature exhibits a parabolic maximum in the center, where the local distribution is maximally mixed.

\subsection{From unitary to dissipative entropy current}
Given the temperatures and chemical potentials of the floating thermoelectric probes, the unitary entropy current injected by a single probe may be calculated from Eq.\ \eqref{unitarys} (using the same assumptions and Sommerfeld expansion), 
\begin{equation}
    I_{P_n}^S= \frac{1}{h} \sum_\alpha T_{{P_n}\alpha} \left( \mu_0 \right)\frac{\pi^2k_B^2}{3}\left(T_{P_n}-T_\alpha \right) \label{singleprobeS},
\end{equation}
and the total entropy production is the sum over all these contributions
\begin{equation}
    \Dot{S}_P = \sum_n I_{P_n}^S \label{spsom}.
\end{equation}
The entropy injected by the probes due to their continuous measurements reflects a constant generation of entanglement between the electrons in the probes and those in the quantum wire, which models the generic inelastic scattering processes in a current carrying wire wherein electronic degrees of freedom in the wire become entangled with environmental degrees of freedom, %
leading to dissipation.

In contrast, in the conventional framework of dissipative transport, %
the entropy production due to Joule heating, %
described by Eq.\ \eqref{jouleh}, takes the form
\begin{equation}
    T_0 \dot{S} =  %
    I_1^{(1)}+I_2^{(1)} = I_1^{(0)}\left( \mu_2-\mu_1\right) = {\cal P}. \label{jheating}
\end{equation} 

Our hypothesis is that the entropy injected into the system by the probes due to their continuous measurements should approach the entropy expected due to Joule heating in the limit that the local thermalization processes introduced by the probes is sufficient to fully relax the outgoing distributions into the reservoirs.  That is, as the number of probes and their coupling strength increases, the ratio of Eqs.\ \eqref{spsom} and \eqref{jheating} should approach unity
\begin{equation}
    \lim_{N\gamma_p\rightarrow \infty} \frac{T_0 \dot{S}_P}{\mathcal{P}} \rightarrow 1. \label{ratio}
\end{equation}
Conventionally \cite{buttiker1,Imry1,bergnano2009,baranger93}, the equilibration processes introduced explicitly in our model by the floating thermoelectric probes are implicitly assumed to take place within the reservoirs themselves. The role of coherence in transport in this context aligns with the discussion given in Ref.\ \cite{buttiker85} (see appendix \ref{apx:oneandtwo}).  %

\begin{figure}[h]	
    \includegraphics[width=1\linewidth]{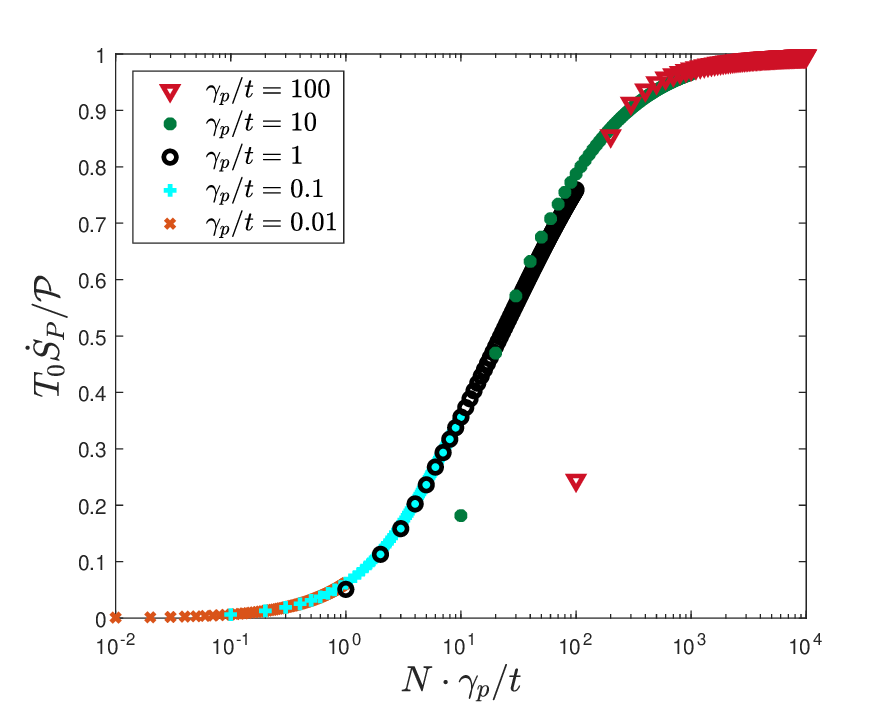}
	\caption{Ratio of total entropy injected by $N$ floating thermoelectric probes (due to their continuous temperature/voltage measurements) to the entropy production expected due to Joule heating in the wire, plotted versus $N\gamma_p/t$,  
    for varying $N \in \{1,\dots ,100\}$ and several values of the probe coupling $\gamma_p$, where $t=2.7$ eV is the hopping matrix element in the quantum wire, $T_0=232\mbox{K}$ is the common temperature of the source and drain electrodes, and the electric bias is $\Delta \mu= 10k_BT_0 =0.2$ eV.}\label{fig:sgen}  
\end{figure}
Direct numerical verification of Eq.\ \eqref{ratio} is presented in Fig.\ \ref{fig:sgen}, which shows moreover 
that this ratio exhibits joint scaling, that is, many weakly coupled probes are as effective as few strongly coupled probes for the purpose of generating entropy. This fact may be of use in future analyses, as it is cheaper to simulate a few rather than many probes. However, there is a limit to this freedom, that is, a single probe, however strongly coupled to the system, can lead to the ratio Eq.\ \eqref{ratio} of at most 1/2 (see appendix \ref{apx:singlelimit}), as indicated by the leftmost points of each data set in 
Fig.\ \ref{fig:sgen}, which represent the results for a single probe.
It is notable that according to Fig.\ \ref{fig:sgen}, a very great number of inelastic scattering events are needed to completely thermalize the outgoing distributions and recover the full entropy due to Joule heating.

\section{$1/N$ scaling: end effects} \label{sec: 1_N}

\begin{figure}[h]
    \begin{center}
        \includegraphics[width=1\linewidth]{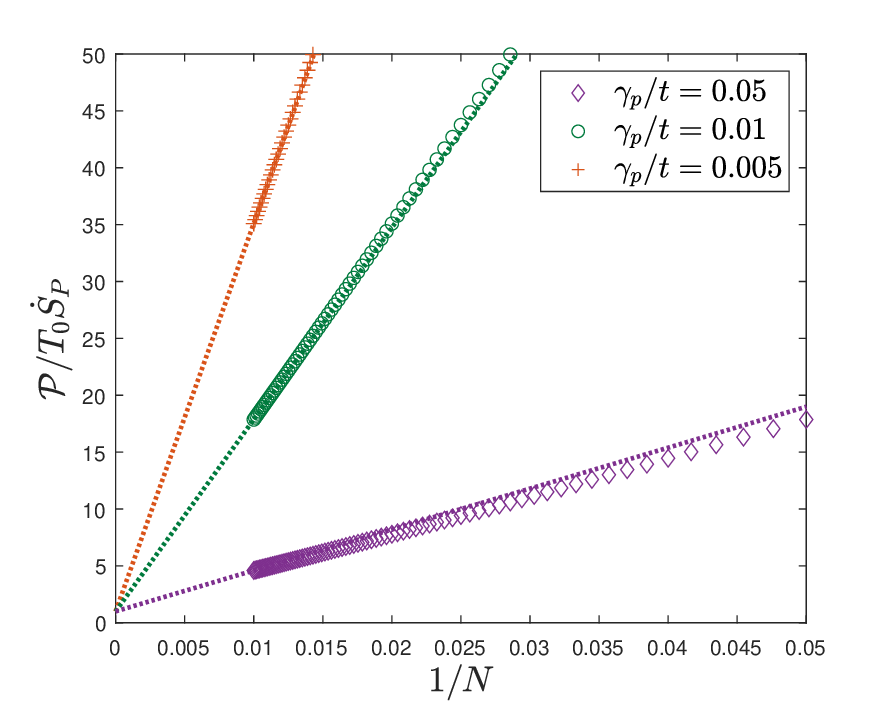}    
    \end{center}
	\caption{Linear character of the ratio of entropy flows for an infinite chain with $N$ probes and multiple $\gamma_p/t$. The dotted lines are defined as the lines that cross the intercept at 1 and the nearest corresponding data point. $T_0=100$K, $t=2.7$eV, $\Delta \mu= 0.1$ eV.}\label{fig:sgen2} 
\end{figure}

\begin{figure*}	
    \includegraphics[width=0.8\linewidth]{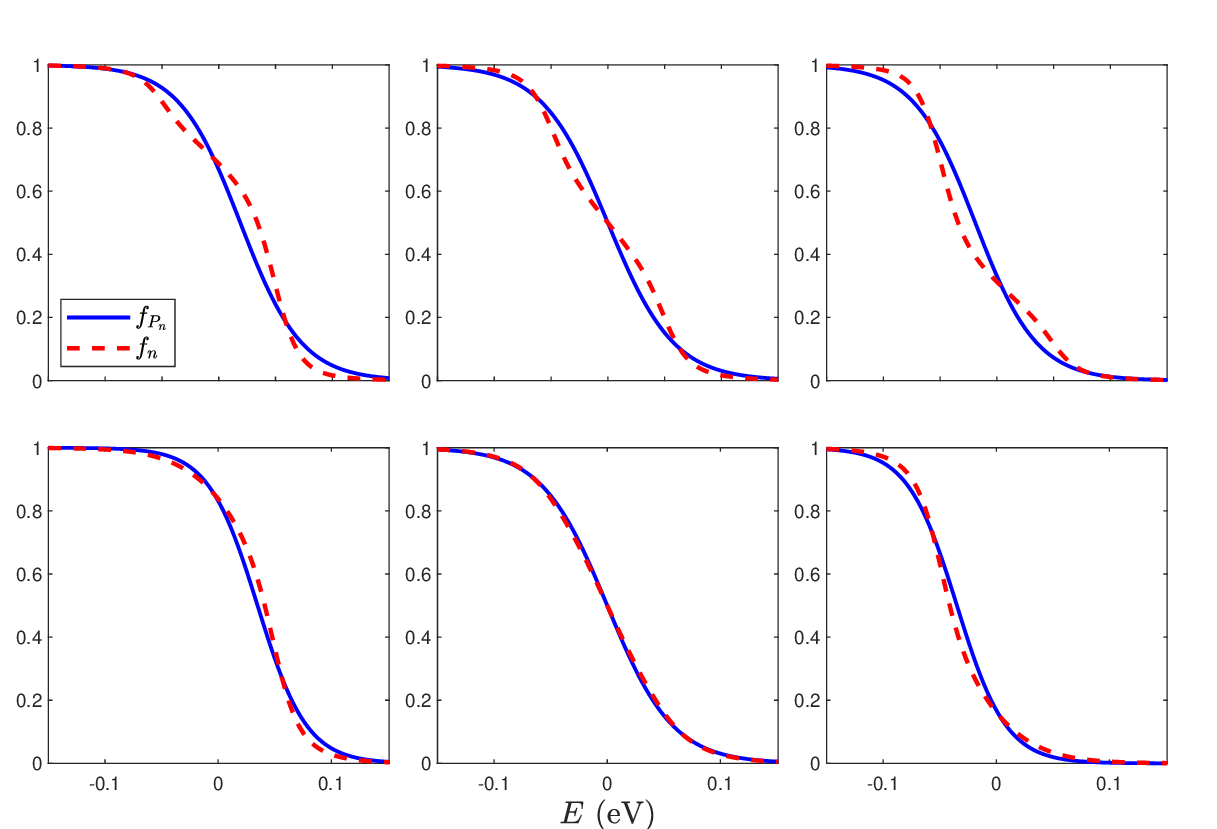}
	\caption{Equilibrium ($f_{P_n}$) and local non-equilibrium ($f_n$) distributions of particle occupations in the sites of an infinite chain as locally measured by $N=11$ probes. The top row corresponds to $\gamma_p/t=0.25$ and the bottom one to $\gamma_p/t=1$. The first column shows the leftmost probe ($N=1$) measurement, the center column the middle of the set of probes ($N=6$) and the third column the rightmost probe ($N=11$). Here $T_0=115$K, $t=2.7$eV, $\Delta \mu=10k_BT_0\approx 0.1$ eV.}\label{fig:fs}  
\end{figure*}

The deficit of entropy injected by the microscopic probe measurements relative to that due to macroscopic Joule heating  
scales as $1/N$ for large $N$ and its slope depends on $\gamma_p/t$ as can be seen in Fig.\ \ref{fig:sgen2}. This fact can be qualitatively understood as an end effect, since the deviation from complete thermalization is greatest at the ends of the wire, as shown in Fig.\ \ref{fig:fs}.

As shown in Fig.\ \ref{fig:mus1}, the local temperature is greatest in the central region of the quantum wire, where thermalization is maximum as the local distribution involves the greatest mixing of the source, drain, and probe distributions.  The local temperatures near the ends of the wire are reduced, reflecting the greater admixture of source and drain distributions, respectively, at temperature $T_0=115\mbox{K}$.  %
This thermalization behavior is confirmed in Fig.\ \ref{fig:fs}, where the equilibrium Fermi-Dirac distributions $f_{P_n}(\epsilon)$ of the probes (characterized by the measured values of $T_{P_n}$ and $\mu_{P_n}$) are contrasted with the local non-equilibrium distributions $f_{n}(\epsilon)$ 
in the quantum wire (see Ref.\ \cite{Shastry2016} and appendix \ref{apx:green} for a definition and discussion of $f_n(\epsilon)$). 
The local distributions in the center of the wire and visibly closer to the corresponding equilibrium probe distributions.
This figure also validates the expressions obtained through Sommerfeld's expansion \eqref{i0ps} and \eqref{i1ps} in the current case, as the local non-equilibrium distributions satisfy the conditions in \cite{Shastry2016}.

 The equilibrium probe and non-equilibrium system distributions differ more the closer they are to the source and drain reservoirs, as equilibration is incomplete near the ends where few scattering events are not enough to fully thermalize the local distribution.  In contrast, near the center of the wire, electrons will have experienced multiple inelastic scattering events such that propagation is akin to a random walk. This is consistent with the fact that a single probe is unable to fully thermalize the local distribution, no matter how strongly it is coupled to the system (see appendix \ref{apx:singlelimit}), and the probes at or near the boundary are mixing a pure Fermi-Dirac distribution from its neighboring source or drain reservoir with the quasi-equilibrium distribution that has propagated down from the other side of the chain. It is also observed that the stronger the coupling to the probes, the more sequential is the transport from one side to the other, the more effective the probes are as scatterers and the less likely particles keep their coherence which results in a local non-equilibrium distribution closer to the equilibrium distribution of the corresponding probes.

\begin{figure}[h]	
    \includegraphics[width=1\linewidth]{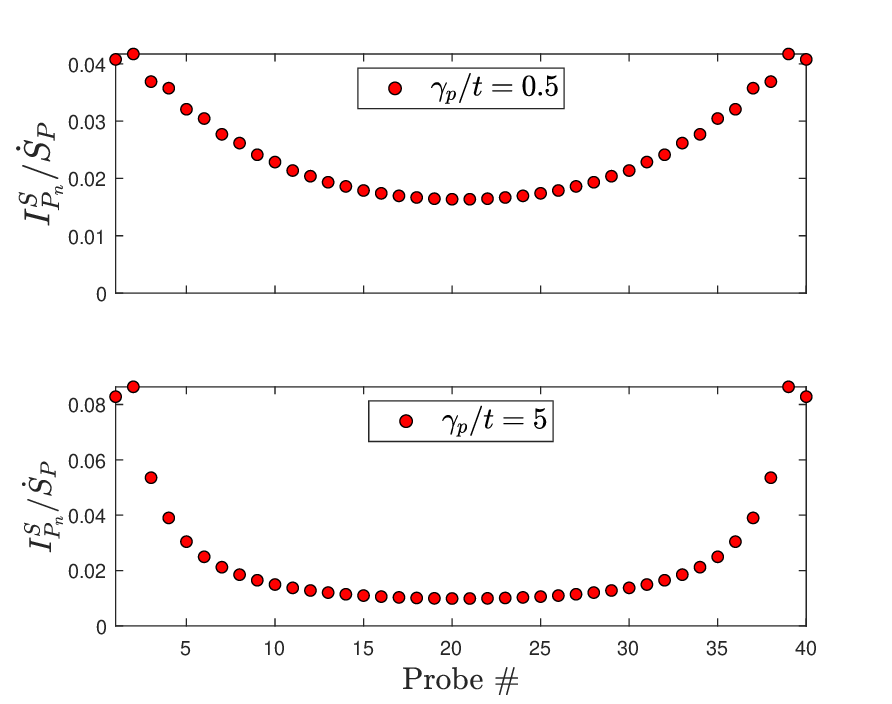}
	\caption{Entropy currents injected by each probe 
    [Eq.\ \eqref{singleprobeS}] normalized by the %
    total entropy injected [Eq.\ \eqref{spsom}]. $T_0=115$K, $t=2.7$eV, $\Delta \mu=10k_BT_0\approx 0.1$ eV in both. }\label{fig:sdif} 
\end{figure}

Fig.\ \ref{fig:sdif} shows that the probes near the ends of the quantum wire inject the greatest amount of entropy due to the less complete thermalization of the distributions at the ends of the wire.  (Note that when a floating probe performs such a measurement on a system that is in thermal and diffusive equilibrium, the steady-state rate of entropy injected by the probe goes to zero.)

The greater deviation from equilibrium of the local distributions near the ends of the wire shown in Fig.\ \ref{fig:fs} can be quantified \cite{shastry2015} by comparing the entropy $S_n$ of the local distribution with the corresponding local entropy $S_{P_n}$ if the system was populated by the equilibrium distribution of its locally coupled probe.  The latter is greater due to the maximum entropy principle, so the difference $\Delta S_n = S_{P_n}-S_n$ can serve as a metric for the distance from local equilibrium, where
\begin{equation}
    S_{P_n} = -\int d\epsilon  \left[ f_{P_n}\ln f_{P_n} + (1-f_{P_n})\ln (1-f_{P_n}) \right] g_{n}
\end{equation}
and 
\begin{equation}
    S_{n} = -\int d\epsilon  \left[ f_{n}\ln f_{n} + (1-f_{n})\ln (1-f_{n}) \right] g_{n},
\end{equation}
where $g_n(\epsilon)=\langle n|A(\epsilon)|n\rangle$ is the local spectrum of the system, $A(\epsilon)$ being the spectral function (see Appendix \ref{apx:green}).
Figure \ref{fig:deltaS} shows that the local distributions are quantitatively much closer to local equilibrium for large probe coupling, as expected, and that the wire ends are significantly farther from local equilibrium than the central regions of the wire, for both weak and strong probe coupling.

\begin{figure}[h]	
    \includegraphics[width=1\linewidth]{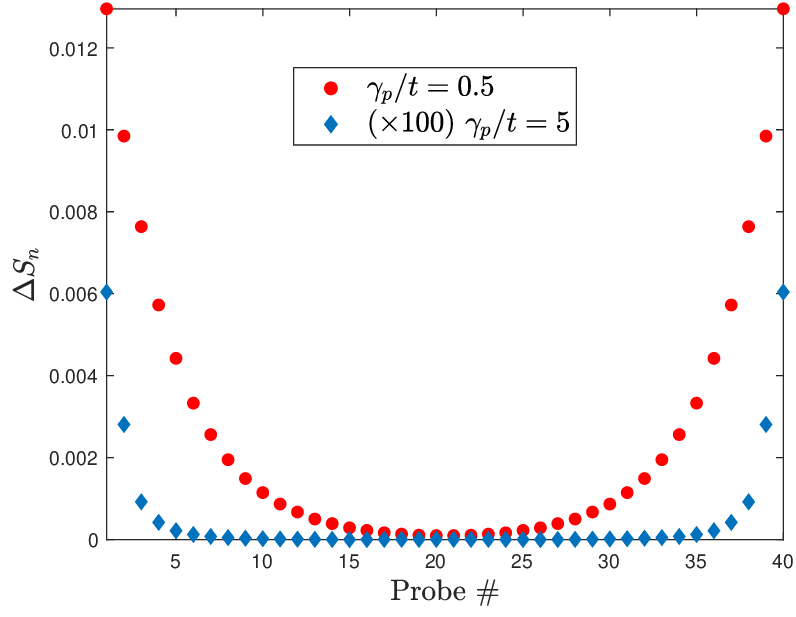}
	\caption{Deviation of the system distribution from local equilibrium for two values of the probe coupling, quantified by the
    difference in entropy $\Delta S_n=S_{P_n}-S_n$ between the equilibrium and non-equilibrium distributions on each site in the chain. The case $\gamma_p/t=5$ is magnified by a factor of 100. $T_0=115$K, $t=2.7$eV, $\Delta \mu=10k_BT_0\approx 0.1$ eV in both cases.}\label{fig:deltaS} 
\end{figure}

\section{Summary} \label{sec:summary}

We have investigated the emergence of irreversibility in an exactly solvable model of a quantum wire under electrical bias, coupled to $N$ floating thermoelectric probes.  Using a new microscopic formula for entropy flow inspired by Refs.\ \cite{marco,evers}, we show how entropy production and irreversibility arise due to the measurement process within a framework of unitary evolution wherein global entropy is conserved.
Here we have succeeded in a project dating back to Boltzmann himself, to derive the 2nd Law of Thermodynamics from a microscopic Hamiltonian.

In our model, the probes serve the role of the ``environment'' of the quantum wire.
The information obtained through the continuous measurements by the probes is %
injected as entropy into the quantum wire, representing entanglement between the electronic degrees of freedom in the probe reservoirs and those in the quantum wire.  We find that as the number of probes and the strength of their coupling to the wire increase, the entropy injected approaches that expected due to Joule heating in a macroscopic analysis of the system.  We have thus derived this macroscopic law from a fully microscopic analysis.

The deficit in entropy production for finite $N$ was shown to scale as $1/N$ due to end effects in the quantum wire.  
The local electron distribution near the ends of the wire remains further from a local equilibrium than the distribution in the central region, where many probes participate in the equilibration process.  The entropy production due to probes near the ends of the wire is thus maximal.

\begin{acknowledgments}
We are grateful to Parth Kumar and Ferdinand Evers for their illuminating discussions and feedback. 
\end{acknowledgments}

\bibliography{entropygen.bib}

\appendix
\section{Green's functions and local non-equilibrium distributions}\label{apx:green}
This appendix is meant to provide the necessary basic information regarding Non Equilibrum Green's Functions (NEGF) methods used in this work, in particular the definition of the retarded and lesser components of the Green's function, and their relation with the spectral function, transmission function, and local distribution. The reader is referred to textbooks \cite{stefanucci,datta} for a more thorough exposition.

The retarded Green's function $G^R(\epsilon)$ describes the coherent evolution of a particle since it's injected into the system until it leaves through a lead or is scattered into a different state \cite{datta}. It can be calculated by solving the Dyson equation 
\begin{equation}
    G^R(\epsilon)=\left( \left[G_0^R(\epsilon)\right]^{-1}-\Sigma^R(\epsilon) \right)^{-1},
\end{equation}
which is usually a matrix equation, and where $G_0^R$ corresponds to the Green's function of the isolated system, and $\Sigma^R$ is the retarded self-energy, used to describe interparticle interactions and the reservoir-system coupling (tunneling self-energy). In both steady-state and equilibrium, the advanced component of the Green's Function  is $G^A(\epsilon)=\left[G^R(\epsilon)\right]^\dagger$. Interparticle interactions are not considered in this work, and in the broadband limit, $\Sigma^R=-\frac{i}{2}\sum_\alpha \Gamma^\alpha$, where the tunneling-width matrices $\Gamma^\alpha$ describe the coupling of the reservoirs and probes to the system, and are independent of energy.

The Green's function of the isolated system satisfies 
\begin{equation}
    \left( \epsilon \mathbb{1}-H_{\rm S} +i \delta^+\right)G_0^R(\epsilon)=\mathbb{1},
\end{equation}
where $\epsilon$ is an energy variable energy, $\delta^+$ an infinitesimal positive constant (required to ensure causality) and $H_{\rm S}$ is the system Hamiltonian matrix on the single-particle Hilbert space, corresponding to \eqref{sysh}.

The transmission function from reservoir $\beta$ to $\alpha$ $T_{\alpha\beta}(\epsilon)$ can be obtained through the use of Green's functions
\begin{equation}
    T_{\alpha\beta}(\epsilon)= \text{Tr}\left[ \Gamma^{\alpha}(\epsilon)G^R(\epsilon)\Gamma^{\beta}(\epsilon)G^A(\epsilon) \right].
\end{equation}

The lesser component of the Green's function in the energy domain $G^<(\epsilon)$ encodes energy-resolved quantum correlations between occupied particle states, it can be used to generalize the concept of distribution functions beyond equilibrium.

This component can be calculated by solving the Keldysh equation
\begin{equation}
    G^<(\epsilon) = G^R(\epsilon) \Sigma^<(\epsilon) G^A(\epsilon),
\end{equation}
where the lesser component of the self-energy $\Sigma^ <$ describes the non-equilibrium scattering processes that insert or remove electrons from a quantum system, acting as a generalized ``in-scattering" term that drives the system away from equilibrium. In the present case, without the presence of interparticle interactions or other memory or time-dependent effects
\begin{equation}
    \Sigma^<(\epsilon) = \sum_\alpha \Gamma^\alpha f_\alpha(\epsilon),
\end{equation}
where $f_\alpha$ is the Fermi-Dirac distribution function of reservoir $\alpha$.

The spectral function $A(\epsilon)$ generalizes the concept of density of states and helps us determine the probability that and added/removed particle in the system has energy $\epsilon$ and their lifetime broadening due to interactions or coupling to the environment. It can be obtained through Green's functions
\begin{equation}
    A(\epsilon) = \frac{i}{2\pi}\left[ G^R(\epsilon)-G^A(\epsilon) \right].
\end{equation}

The spectral function along with the lesser component of the Green's function of a system in equilibrium can be used to obtain the equilibrium distribution function of the system
\begin{equation}
        G_{\rm eq}^<(\epsilon) = 2\pi i A(\epsilon) f_{\rm eq}(\epsilon),
\end{equation}
which serves as motivation for defining a local non-equilibrium distribution function of the system (one that is generically out of equilibrium), as sampled by a probe $P_n$, $f_{n}(\epsilon)$ discussed in \cite{Shastry2016,Stafford2016} and defined as
\begin{equation}
    f_{n}(\epsilon) = \frac{\Tr \left\{ \Gamma^{P_n}G^<(\epsilon) \right\}  }{ 2\pi i \Tr \left\{ \Gamma^{P_n}A(\epsilon) \right\}}.
\end{equation}

\section{Entropy generation limit for a single probe}\label{apx:singlelimit}

Considering \eqref{unitarys} in the case of a single probe $P$ and the two reservoirs 1 and 2 and
due to \eqref{noSprod}, with a minus sign to highlight the fact that this is the unitary entropy production rate injected into the system by the probe
\begin{equation}
    -I_P^S = \frac{1}{h}\sum_\beta\int d\epsilon\  T_{P\beta}(\epsilon) \left[ s_P(\epsilon)-s_\beta(\epsilon) \right]
\end{equation}
Condition \eqref{probecond} for a single probe
\begin{equation}
    I_P^{(\nu)} =\frac{1}{h} \sum_\beta \int d\epsilon\ \left( \epsilon-\mu_P \right)^\nu T_{P\beta}(\epsilon)\left[
     f_\beta-f_P\right] = 0,  \label{condPapx}
\end{equation}
for $\nu=0,1$. In the transport regime considered in this work, $T_{P1}=$const., $T_{P2}=$const., and accounting \eqref{condPapx} fixes both $\mu_P$ and $T_P$
\begin{equation}
    \mu_P=\frac{T_{P1}\mu_1+T_{P2}\mu_2}{T_{P1}+T_{P2}},
\end{equation}
and
\begin{equation}
    \frac{\pi^2k_B^2}{3}T^2_P = \frac{\pi^2k_B^2}{3}T^2_0 + \frac{T_{P1}T_{P2}}{\left( T_{P1}+T_{P2} \right)^2}(\mu_1-\mu_2)^2
\end{equation}
Considering the same regime leading to the validity of the Sommerfeld expansion as in \eqref{singleprobeS}
\begin{equation}
    -I_P^S \approx \frac{1}{h} (T_{P1}+T_{P2}) \frac{\pi^2k_B^2}{3}(T_P-T_0)
\end{equation}
Meaning that the total entropy production rate in the reservoirs is
\begin{eqnarray}
     I_1^S+I_2^S  & \approx & \frac{1}{2} \frac{1}{h} \frac{T_{P1}T_{P2}}{T_{P1}+T_{P2}} \frac{(\mu_1-\mu_2)^2}{T_0}. \label{onePSuni}
\end{eqnarray}
In contrast, the total entropy production rate due to the process of Joule heating according to conventional thermodynamics can be obtained by considering
\begin{eqnarray}
    \sum_\alpha I_\alpha^{(1)} &=& \sum_\alpha I_\alpha^E-\sum_\alpha \mu_\alpha I_\alpha^{(0)}\\
    &=& -\sum_\alpha \mu_\alpha I_\alpha^{(0)} - (\mu_1-\mu_2)I_1^{(0)}
\end{eqnarray}
The total particle current across the chain is 
\begin{equation}
    I_1^{(0)} = \frac{1}{h}\left( T_{12} + \frac{T_{1P}T_{P2}}{T_{P1}+T_{P2}} \right) (\mu_2-\mu_1)
\end{equation}
Assuming a full conversion of electrical energy into heat (Joule heating)
\begin{equation}
    I^Q_1+I^Q_2 = \frac{1}{h}\left( T_{12} + \frac{T_{1P}T_{P2}}{T_{P1}+T_{P2}} \right) (\mu_2-\mu_1)^2
\end{equation}
Then the entropy production rate due to Joule heating is
\begin{equation}
    \frac{I^Q_1+I^Q_2}{T_0} = \frac{1}{h}\left( T_{12} + \frac{T_{1P}T_{P2}}{T_{P1}+T_{P2}} \right) \frac{(\mu_2-\mu_1)^2}{T_0}. \label{onePSJ}
\end{equation}
Since $T_{12} \geq 0$, reaching the equality in the limit of very strong coupling $\gamma_p/t\rightarrow\infty$, meaning no significant probability of direct tunneling from one reservoir to the other, the entropy production \eqref{onePSuni} is always less than or equal to half that of \eqref{onePSJ} and can never match it, regardless of how strongly the probe is coupled to the system. 

This is an important manifestation of the distinction between equilibration and decoherence. Even a completely inelastic transport is not sufficient to saturate the entropy production due to Joule heating. Though related, complete decoherence of the electron flow does not equate to complete equilibration, because decoherence refers to electron phases and equilibration refers to distribution in energy, it is not possible to have decoherence without inelastic scattering which causes relaxation in energy, but maximizing one does not guarantee the other be a maximum as well.

\section{Linear and quadratic dependence of total resistance}\label{apx:oneandtwo}

It is particularly simple to describe the total resistance in the limits of weak and strong couplig $\gamma_p/t$. For strong coupling, it is assumed that transport is sequential, meaning that particles have a high probability of transporting into their nearest-neighbor site and negligible probability of going into others. 

In a sequential transport regime, resistances are related to transmission probabilities as $R = \frac{h}{e^2}\sum_i\frac{1}{T_i}$, where spin does not play any role in the current model and with $T_i$ the transmission probability into different channels which can be expressed in terms of Green's functions.

\subsection{Weak coupling limit}

Considering a single probe, from Eq.\ \eqref{i0ps} and applying the condition \eqref{probecond},
\begin{equation}
    \mu_P=\frac{T_{P1}\mu_1+T_{P2}\mu_2}{T_{P1}+T_{P2}}
\end{equation}
which can be used to describe the overall current,
\begin{equation}
    I_{1}^{(0)} = \frac{1}{h}\left[ T_{12}+ \frac{T_{1P}T_{P2}}{T_{P1}+T_{P2}} \right] (\mu_P- \mu_1).
\end{equation}
The transmission functions can be calculated with NEGF (see \ref{apx:green}). Due to the time-reversal symmetry, $T_{\alpha \beta} = T_{\beta \alpha}$. It follows
\begin{equation}
    T_{\alpha\beta}(\epsilon) = \frac{\Gamma^\alpha\Gamma^\beta}{(\epsilon-\varepsilon)^2+\overline{\Gamma}^2},
\end{equation}
with $\Gamma^P = \gamma_p$, $\Gamma^1 = \Gamma^2 = 2t$, $\overline{\Gamma} = \frac{1}{2} \sum_\alpha \Gamma^\alpha$. The system's energy is fixed at $\varepsilon=\mu_0=0$, following the assumption of constant transmission, then
\begin{equation}
    T_{12}(\epsilon = \varepsilon) = \frac{4t^2}{\overline{\Gamma}^2},
\end{equation}
is the transmission of direct tunneling from 1 to 2, whereas
\begin{equation}
    T_{1P} = \frac{2t\gamma_p}{\overline{\Gamma}^2} \approx \frac{\gamma_p}{2t}  = T_{2P},
\end{equation}
by keeping only the leading order in $\gamma_p$, and they represent the transmission functions between the reservoirs and the probe. It follows that
\begin{equation}
     I_{1}^{(0)} = \frac{1}{h}\left[ 1-\frac{\gamma_p}{4t} \right] (\mu_P- \mu_1).
\end{equation}
The total effective transmission probability in this case would be
\begin{eqnarray}
    \tilde{T}_{12} &=& T_{12} + \frac{T_{1P}T_{P2}}{T_{P1}+T_{P2}} \\
    &=& \frac{1}{1+\frac{\gamma_p}{4t}} \approx 1-\frac{\gamma_p}{4t},
\end{eqnarray}
following the small $\gamma_p/t$ assumption. Consequently 
\begin{equation}
    R = \frac{h}{e^2} \frac{1}{\tilde{T}_{12}} =  \frac{h}{e^2} \left( 1 + \frac{1}{4}\frac{\gamma_p}{t} \right),
\end{equation}
showing a linear dependence with $\gamma_p/t$.

For more than one probe, it is important to notice that contributions to the total effective transmission $\tilde{T}_{12}$ due to the introduction of probe-probe transmissions are of higher order
\begin{equation}
    T_{P_iP_j} = \left( \frac{\gamma_p}{2t} \right)^2.
\end{equation}
Ultimately, it can be inductively shown that for $N$ weakly coupled probes, their contribution to the total transmission is additive*
\begin{equation}
    \tilde{T}_{12} \approx 1-N\frac{\gamma_p}{4t},
\end{equation}
which results in
\begin{equation}
    R = \frac{h}{e^2} \left( 1 + N\frac{1}{4}\frac{\gamma_p}{t} \right) + \mathcal{O}\left( \frac{\gamma_p}{t} \right)^2.
\end{equation}

\begin{figure}[h]	
    \includegraphics[width=1\linewidth]{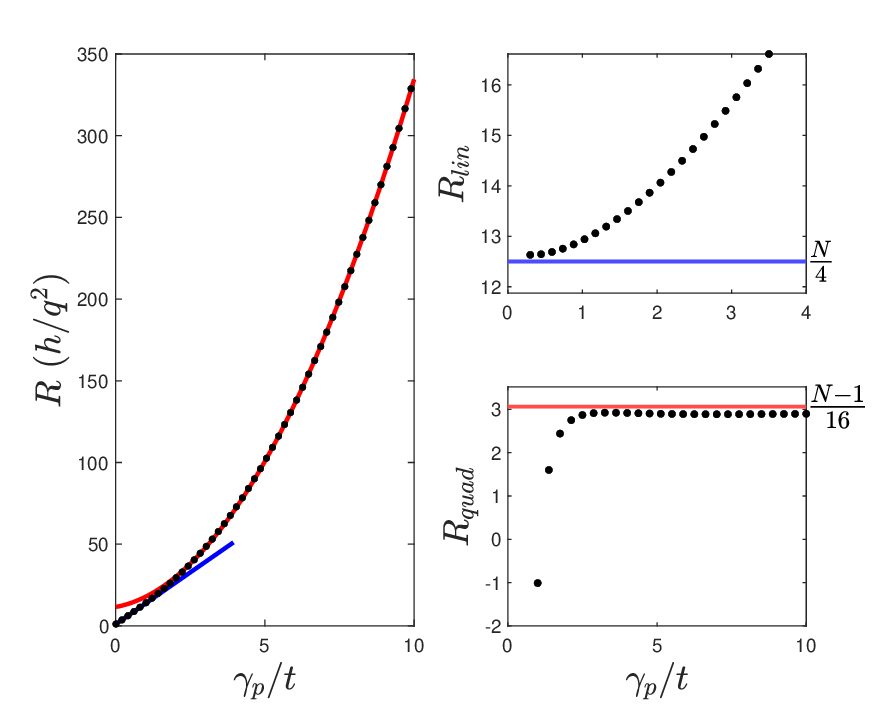}
	\caption{Left: Total resistance in the sequential transport regime from transmission probabilities in an infinite chain with 50 probes. Top right: $R_{lin}$ is the best-fit bias term subtracted from the total resistance divided by $\gamma_p/t$ to isolate the linear dependence. Bottom right: $R_{quad}$ is the best-fit linear and bias terms subtracted from the total resistance divided by $(\gamma_p/t)^2$ to isolate the quadratic dependence.
    Linear and quadratic dependence in both limits are consistent with the analysis in \ref{apx:oneandtwo}. }\label{fig:totres}  
\end{figure}

A verification of the scaling of the total resistance in this regime is shown in Fig.\ \ref{fig:totres}

\subsection{Strong coupling limit} 
For multiple probes, a general expression for total resistance can be attained by assuming the transmission function among two adjacent probes is the same, but its Green's function are easily obtainable for two probes. 

Considering sequential transport and modeling each of the sites as a classical resistance contributing additively to the total resistance
\begin{equation}
    R=R_c+(N-1)R_{nn},    
\end{equation}
 where $R_c$ represents the contact resistance and $R_{nn}$ the resistance associated to the rest of the $N-1$ links (assumed identical) forming the chain, related to the transmission function among neighboring probes
 \begin{equation}
  T_{nn} = \Gamma_P^2 \vert G_{nn}^R\vert ^2.   
 \end{equation}

In the two probe case, the retarded Green's function takes the form 
\begin{equation}
    G^R = \frac{1}{t^2_0+\left( t_0+\frac{\gamma_p}{2} \right)^2}
    \begin{bmatrix}
        i\left( t_0+\frac{\gamma_p}{2} \right) & t_0 \\
        t_0 & i\left( t_0+\frac{\gamma_p}{2} \right)
    \end{bmatrix},
\end{equation}
then
 \begin{equation}
     T_{nn} = \frac{\gamma_p^2 t_0^2}{t_0^2+\left( t_0+\frac{\gamma_p}{2} \right)^2} = \frac{\gamma_p^2}{t_0^2} \frac{1}{\left( 1+ \left( 1+\frac{\gamma_p}{2t_0} \right)^2 \right)^2},
 \end{equation}
in the limit of large $\gamma_p/t$ this simplifies to
\begin{equation}
    T_{nn} \xrightarrow[\frac{\gamma_p}{t_0} \to \infty]{} \frac{16t_0^2}{\gamma_p^2},
\end{equation}
which yields a total resistance that depends quadratically on $\gamma_p/t$
\begin{equation}
    R = R_c+(N-1)\frac{h}{e^2} \frac{\gamma_p^2}{16 t_0^2}.
\end{equation}

This series resistance model is of qualitative use, as it helps us understand the general trends and confirm some of the intuitions about the system's resistance, even though it lacks complete quantitative predictive power.

\end{document}